\documentclass[12pt]{article}
\pdfoutput=1

\usepackage[pdftex]{graphicx}
\usepackage{lscape} 
\usepackage{array}
\usepackage{extarrows}

\usepackage{fix-cm}

\usepackage[makeroom]{cancel}
\usepackage[normalem]{ulem}
\usepackage{amsmath,amssymb}
\usepackage{slashed}
\usepackage{xcolor}
\usepackage{cite}
\usepackage{pdflscape}
\usepackage{multirow}
\usepackage[thinlines]{easytable}
\usepackage{url}
\usepackage[utf8]{inputenc}
\usepackage{longtable}
\usepackage{booktabs}

\newcommand{\eq}[1]{\begin{equation} #1 \end{equation}}
\newcommand{\eqa}[1]{\begin{eqnarray} #1 \end{eqnarray}}

\newcommand{\GeV}{\,{\rm GeV}}

\newcommand{\Eq}[1]{Eq.~(\ref{#1})}
\newcommand{\REF}[1]{Ref.~\cite{#1}}
\newcommand{\Sec}[1]{Section~\ref{#1}}

\newcommand{\cW}{{\cal W}}
\newcommand{\cO}{{\cal O}}
\newcommand{\B}{{\cal B}}
\newcommand{\A}{{\cal A}}

\newcommand{\av}[1]{\langle #1 \rangle}

\newcommand{\re}{{\rm Re}}
\newcommand{\im}{{\rm Im}}

\newcommand{\V}{{\cal V}}

\newcommand{\Leff}{{\cal L}_\text{eff}}
\newcommand{\sla}[1]{#1 \!\!\!/}

\begin{document}


\begin{titlepage}

\vspace*{-2cm}
\begin{flushright}
SI-HEP-2020-18\\
P3H-20-037\\
TUM-HEP-1272/20
\end{flushright}

\vspace{0.7cm}

\begin{center}
\bf
\fontsize{19.6}{24}\selectfont
Three-Body Non-Leptonic Heavy-to-heavy \\
$B$ Decays at NNLO in QCD
\end{center}

\vspace{0cm}

\begin{center}
\renewcommand{\thefootnote}{\fnsymbol{footnote}}
{Tobias Huber$^a$, Javier Virto$^{b}$ and K.~Keri Vos$^c$}
\renewcommand{\thefootnote}{\arabic{footnote}}
\setcounter{footnote}{0}

\vspace*{.8cm}
\centerline{${}^a$\it  Naturwissenschaftlich-Technische Fakult\"at, Universit\"at Siegen,}
\centerline{\it Walter-Flex-Stra{\ss}e 3, 57068 Siegen, Germany}
\vspace{2.5mm}

\centerline{${}^b$\it Departament de F\'isica Qu\`antica i Astrof\'isica, Institut de Ci\`encies del Cosmos,}
\centerline{\it Universitat de Barcelona, Mart\'i Franqu\`es 1, E08028 Barcelona, Catalunya}\vspace{1.3mm}

\centerline{${}^c$\it Physics Department T31, Technische Universit\"at M\"unchen,} \centerline{\it James Franck-Stra{\ss}e 1, D-85748 Garching, Germany}
\vspace{1.3mm}

\vspace*{.2cm}

\end{center}

\vspace*{10mm}
\begin{abstract}\noindent\normalsize
Exclusive non-leptonic two-body decays of $B$ mesons have been studied extensively in the past two decades within the framework of factorization. However, the exploration of the corresponding three-body case has only started recently, in part motivated by new data.
We consider here the simplest non-leptonic three-body $B$ decays from the point of view of factorization, namely heavy-to-heavy transitions. We provide a careful derivation of the SCET/QCDF factorized amplitudes to NNLO in $\alpha_s$, and discuss the numerical impact of NLO and NNLO corrections. We then study the narrow-width limit,  showing that the three-body amplitude reproduces analytically the known quasi-two-body decay amplitudes, and compute finite-width corrections. Finally, we discuss certain observables that are sensitive to perturbative NLO and NNLO corrections and to higher Gegenbauer moments of the dimeson LCDAs. This is the first study of non-leptonic three-body $B$ decays to NNLO in QCD. 

\end{abstract}

\end{titlepage}
\newpage 

\renewcommand{\theequation}{\arabic{section}.\arabic{equation}} 

\setcounter{tocdepth}{2}
\tableofcontents


\section{Introduction}
\label{sec:intro}
\setcounter{equation}{0}

For more than two decades, non-leptonic decays of $B$ mesons have been known to provide unique information on three fronts: CP violation and the angles of the CKM matrix, the structure of QCD in the presence of heavy quarks and energetic light particles, and beyond-the-Standard Model physics in the quark sector.

Experimental information on non-leptonic $B$ decays has been collected progressively over the years, at CLEO, the B-factories BaBar and Belle~\cite{Bevan:2014iga}, the Tevatron, and currently at the LHC, most prominently at LHCb. Future experimental programs at the upgraded LHC and at Belle~II also include serious plans for analyses of non-leptonic $B$ decays~\cite{Bediaga:2018lhg,Kou:2018nap}.

 On the theory side, the calculation of the complicated non-perturbative matrix elements which are present in the amplitude is the bottleneck to precision predictions. The most important leap towards the deconstruction of these matrix elements came with the development of the QCDF/SCET approach~\cite{Beneke:1999br,Beneke:2000ry,Bauer:2000yr,Bauer:2001cu}, which allows to factorize the amplitudes of two-body decays such as $B\to\pi\pi$ or $B\to D\pi$ and express them in terms of local form factors, meson light-cone distribution amplitudes (LCDAs) and perturbative matching coefficients (also known as `hard kernels'), up to corrections of order $\Lambda_{QCD}/m_b$. 
 Perturbative matching coefficients are currently known up to NNLO in QCD~\cite{Beneke:2005vv,Kivel:2006xc,Beneke:2006mk,Pilipp:2007mg,Bell:2007tv,Bell:2009nk,Beneke:2009ek,Kim:2011jm,Huber:2015bva,Huber:2016xod,Bell:2015koa,Bell:2020qus}.

The extension of the QCDF/SCET approach beyond two-body decays (i.e.~three-body), although discussed already early on~\cite{Beneke,Stewart}, has only been pursued relatively recently~\cite{Krankl:2015fha}. 
Contrary to two-body decays, where the phase space merely consists of a single point, the phase space of the three body decay is two dimensional, and can be parameterized by the invariant masses of two pairs of final-state particles. The factorization properties of the three-body decay amplitude in the heavy-quark limit will depend on the scaling of these invariant masses with $m_b$, that is, on the region of phase space~\cite{Krankl:2015fha}. The situation that closely resembles the two-body case arises in the edges of phase space, where two invariant masses are large and the third is small. This includes the resonance region, containing quasi-two-body decays such as $B\to \pi K^*$ or $B\to D \rho$, but understanding these quasi-two-body decays beyond the narrow-width limit requires studying the underlying three-body decay (see e.g.~\cite{Krankl:2015fha,Cheng:2017smj, Bediaga:2017axw,Klein:2017xti,Descotes-Genon:2019bud,Mannel:2020abt,Bediaga:2020ztp}).
The factorization of the three-body amplitude $B\to M M_1 M_2$ in the region where the invariant mass of the pair $(M_1M_2)$ is small is in fact virtually identical to that of the two-body decay, the only difference arising in some of the form factors and LCDAs appearing in the factorization formula, which must be generalized to $B\to M_1M_2$ form factors and dimeson LCDAs. The hard matching coefficients, on the other hand, are identical and one can use the NNLO calculations available from two-body decays. 
The generalized form factors appear in other classes of decays such as $B\to \pi\pi\ell\nu$~\cite{Faller:2013dwa} or $B\to K\pi \ell\ell$~\cite{Descotes-Genon:2019bud}, and thus significant efforts have been devoted to their calculation~\cite{Hambrock:2015aor,Cheng:2017smj,Cheng:2017sfk,Descotes-Genon:2019bud}.
Dimeson LCDAs are, however, not very well-known (see e.g.~\cite{Polyakov:1998ze}), and it is one of the main purposes of the present paper to further investigate their role in three-body non-leptonic $B$ decays.

While this framework has been already put to work phenomenologically in decays to light mesons such as $B\to\pi\pi\pi$ (see e.g.~\cite{Klein:2017xti}), it seems clear that in order to deepen our insight on the theory of factorization in three-body non-leptonic decays one should study in detail the simplest of the cases.
These are the heavy-to-heavy non-leptonic decays $B\to D M_1M_2$ where the flavor of the $B$-meson spectator is picked up by the $D$ meson, in the region of low invariant mass of the $(M_1M_2)$ system.
To leading power in $\Lambda_{QCD}/m_b$ these amplitudes are factorized into well-known $B\to D$ form factors, matching coefficients (known to NNLO in $\alpha_s$~\cite{Huber:2015bva,Huber:2016xod}), and dimeson $M_1M_2$ LCDAs\,; schematically,
\eq{
\A(\bar B^0\to D^+ K^- \pi^0) \sim F^{B\to D} \int du\ T(u,\mu)\ \phi_{K\pi}(u,\mu) + \cO(\Lambda_{QCD}/m_b)\ .
}
This amplitude is under more theoretical control than other three-body decays to light mesons where there are two more terms depending on generalized $B\to M_1M_2$ form factors and the $B$-meson LCDA.
In this paper we study the heavy-to-heavy three-body decays such as the one above in the region of low invariant mass of the light dimeson system. We derive the factorization of the matrix elements, lay out the structure of the amplitudes at NNLO, and explore phenomenological consequences of these results.

The structure of the paper is the following. We start in~\Sec{sec:kinematics} with a general description of the kinematics of the three-body decay. In~\Sec{sec:QCDF} we derive the SCET/QCDF factorization formula for the matrix elements at NNLO, discussing also the relevant non-perturbative quantities that appear in the factorization formula (form factors and dimeson LCDAs), and study the numerical size of NLO and NNLO perturbative corrections.
In~\Sec{sec:nwl} we consider the narrow-width limit and relate in a precise manner the three-body decays to the quasi-two body $B\to D\rho$ and $B\to D K^*$ decays discussed extensively in the literature, and we compute the relevant corrections to this limit.
\Sec{sec:higherorderQCDF} contains a discussion on the effect of higher order corrections on a class of observables built of ratios of bins in the Dalitz plot, both for the $D\pi\pi$ and the $D K\pi$ case.
We conclude in~\Sec{sec:conclusions}.

\section{Kinematics and phase space of the three-body decay}
\label{sec:kinematics}
\setcounter{equation}{0}

We consider the decays $\bar B^0(p)\to D^+(q)\,M^-(k_1)\pi^0(k_2)$ with $M=\pi, K$.
In the $B$-meson rest frame, and choosing $\vec k=\vec k_1 + \vec k_2$ in the $+\hat z$ direction, we have:
\eqa{
p^\mu &=& m_B\,v^\mu = \frac{n^\mu}2 m_B + \frac{\bar n^\mu}2 m_B\ ;\\
k^\mu &\equiv& k_1^\mu + k_2^\mu =  \frac{n^\mu}2 k^- + \frac{\bar n^\mu}2 k^+\ ;\\
q^\mu &=& m_D\,v^{\prime \mu} = \frac{n^\mu}2 (m_B-k^-) + \frac{\bar n^\mu}2 (m_B-k^+)\ ;\\
\overline k^\mu &\equiv& \bigg(1-\frac{\Delta m_{M\pi}^2}{k^2}\bigg) k_1^\mu - \bigg(1+\frac{\Delta m_{M\pi}^2}{k^2}\bigg) k_2^\mu = \frac{n^\mu}2 \bar k^- + \frac{\bar n^\mu}2 \overline k^+ + \overline k_\perp^\mu\ ,
}
where $v,v'$ are the velocities of the $B$ and $D$ mesons respectively.
The light-cone vectors $n,\bar n$ are given in this frame by $n=(1,0,0,1)$ and $\bar n=(1,0,0,-1)$, and $\Delta m_{M\pi}^2 \equiv k_1^2-k_2^2 = m_M^2-m_\pi^2$, such that $k\cdot\overline k=0$.
In full generality,
\eqa{
k^{\pm} &=& \frac{m_B^2-m_D^2+k^2 \mp \sqrt{\lambda_{BD}(k^2)}}{2m_B}\ ,\\
\overline k^{\pm} &=& \mp \frac{k^\pm \sqrt{\lambda_{M\pi}(k^2)} \cos\theta_\pi}{k^2}\ ,
}
in terms of kinematic K\"all\'en functions $\lambda_{ij}(k^2) = (m_i^2-m_j^2)^2 + k^4 - 2(m_i^2+m_j^2)k^2$.
Here  $\theta_\pi$ is defined as the angle between the three-momenta of the neutral pion~($\vec k_2$) and the $B$-meson~($\vec p$) in the $(M\pi)$ rest frame, in which $\vec k=0$ holds. This defines all momenta in terms of the two kinematic variables $(k^2,\theta_\pi)$, which parameterize the phase space. 

The decay amplitudes are therefore functions of $(k^2,\theta_\pi)$. The dependence on these two variables can be factorized by expanding the amplitudes in partial waves,
\eq{
\A(k^2,\theta_\pi) = \sum_{\ell=0}^\infty \A^{(\ell)}(k^2) P_\ell(\cos\theta_\pi)\ ,
\label{eq:APW}
}
where $P_\ell(x)$ are Legendre polynomials: $P_0(x)=1$, $P_1(x)=x$, etc. In the case $M\pi=\pi^-\pi^0$, the dimeson is an isospin-one state, and therefore the sum runs over odd $\ell$ only. In the case $K^-\pi^0$, even partial waves can contribute significantly, e.g.\ resonance contributions such as $\bar B^0\to D^+ K_0^{*-}(800)$ will contribute to the $S$-wave amplitude $\A^{(0)}$.
It is worth noting that the partial wave expansion cannot converge for values of $k^2$ where cross-channel resonance contributions such as $B\to D^{*+}M^-$ are relevant. This issue may be addressed by the isobar method, for example, a subject we will not comment on any further (see e.g.~\cite{Virto:2016fbw,Guo:2014vya}).

The differential decay rate in terms of the two phase-space variables $(k^2,\theta_\pi)$ is given by
\eq{
\frac{d\Gamma(\bar B\to D^+M^-\pi^0)}{dk^2\,d\cos\theta_\pi}
= \frac{\sqrt{\lambda_{BD}(k^2)\,\lambda_{M\pi}(k^2)}}{64 (2\pi)^ 3 k^2 m_B^3} |\A(\bar B\to D^+M^-\pi^0)|^2\ ,
\label{eq:dGamma}
}
for $M=\{\pi,K\}$. For the quasi-two-body decay rates into ``stable'' vector mesons $V=\{\rho,K^*,\dots\}$ we have, on the other hand,
\eq{
\Gamma(\bar B\to D^+V^-) = \frac{\sqrt{\lambda_{BD}(m_V^2)}}{16\pi m_B^3}|\A(\bar B\to D^+V^-)|^2 \, ,
}
which will be useful when checking the narrow-width limit.

\section{Decay amplitudes in QCD factorization at NNLO}
\label{sec:QCDF}
\setcounter{equation}{0}

\subsection{Setup}

The $\bar B^0\to D^+ L^-$ amplitudes, with $L^-=\{\pi^-,\rho^-,\pi^-\pi^0,K^-,K^{*-},K^-\pi^0,\dots \}$ a light hadronic state, are mediated by $b\to c \bar u d$ and $b\to c \bar u s$ effective operators in the effective weak Lagrangian. In the Standard Model, only two dimension-six operators are relevant\,\footnote{For the full set of dimension-six $b\to c \bar u x$ operators beyond the Standard Model see e.g.~\REF{Aebischer:2017gaw}.},
\eq{
\Leff^{(6)} = -\frac{4 G_F}{\sqrt2} \,V_{ux}^* V_{cb} \, (C_1 Q_1 + C_2 Q_2) + h.c.
}
with $x=d,s$ and\,\footnote{Note that these operators differ by a factor of four from those in \REF{Huber:2016xod}.}
\eq{
Q_1 = (\bar c \gamma^\mu P_L T^a b)\,(\bar x \gamma_\mu P_L T^a u) \ ,
\quad Q_2 = (\bar c \gamma^\mu P_L b)\,(\bar x \gamma_\mu P_L u) \ .
}
The cases $x=s,d$ apply to final states with and without strangeness, respectively, and thus we will not distinguish between both sets of operators.
The decay amplitudes are then given by
\eq{
\A(\bar B^0\to D^+ L^-) = \frac{4 G_F}{\sqrt2} \,V_{ux}^* V_{cb}
\,\big[ C_1 \av{Q_1} + C_2 \av{Q_2} \big]
}
with
\eq{
\av{Q_i} \equiv \av{D^+ L^- | Q_i | \bar B^0}\ .
}
We are mostly interested in $L^- = \pi^-\pi^0$ and $L^- = K^-\pi^0$, but will also reproduce the cases $L^-=\pi^-, K^-$ and $L^- = \rho^-,  K^{*-}$ in order to validate the general approach.

Given that the short-distance coefficients $C_i$ are perturbative and well-known by now (see~\cite{Bobeth:1999mk,Gambino:2003zm,Gorbahn:2004my} and references therein), the theoretical challenge is to compute the a priori non-perturbative hadronic matrix elements $\av{Q_i}$. While this is an unsolved problem in general, it is known that if the heavy charmed meson $D^+$ retains the light degrees of freedom from the parent $B$ meson, and if the hadronic state $L^-$ is light (and therefore very energetic in the $B$-meson rest frame), the matrix elements can be treated within the Soft-Collinear Effective Theory (SCET)~\cite{Bauer:2000yr,Bauer:2001yt,Beneke:2002ph,Beneke:2002ni,Becher:2014oda}, where they factorize at the leading order in the EFT counting~\cite{Beneke:2000ry,Bauer:2001cu}.

We thus consider the kinematic region where the pair $(M^-\pi^0)$
has small invariant mass, that is $(k_1+k_2)^2 \equiv k^2 \sim \lambda^2 m_b^2 \ll m_b^2$, where $\lambda = \Lambda_{QCD}/m_b \ll 1$ is the power-counting parameter. From the general kinematic relations in the previous section it follows that $(k^+,k^-)\sim m_b\,(\lambda^2,1)$, and thus the dimeson $(M^-\pi^0)$ has small invariant mass and large energy in the $n$ direction.
In this region the kinematics and the factorization properties are very similar to the two body decay $\bar B^0\to D^+\pi^-$~\cite{Beneke:2000ry,Bauer:2001cu}.

\subsection{SCET matching at NNLO and factorized matrix elements}

We match the QCD operators $Q_i$ onto SCET operators $\cO_k(t)$ via
\eqa{
Q_i &=& \sum_{k} \int d\hat t \, C_{ik}(\hat t)\, \cO_k(t) \, ,
}
where $\hat t  = t \, k^-$ and the SCET operators read
\eqa{
\cO_1(t) &=&
[\bar h_{v'} \sla{n} P_L h_v]
[\bar \chi_n^{(d)}(t\bar n)\frac{\sla{\bar n}}{2} P_L \chi_n^{(u)}(0)] \ ,\\[1mm]
\cO_2(t) &=&
[\bar h_{v'} \sla{n} P_R h_v]
[\bar \chi_n^{(d)}(t\bar n)\frac{\sla{\bar n}}{2} P_L \chi_n^{(u)}(0)]\ .
}
These operators are non-local on the light-cone, and in our notation we are suppressing the Wilson lines required to make the operators gauge invariant.
At leading power, the matrix elements can then be further decomposed as
\eqa{
\av{Q_i} &=& \sum_{k} \int d\hat t \, C_{ik}(\hat t)\, \av{\cO_k(t)} \nonumber\\[2mm]
&=& \int d\hat t \, \Big(
C_{i1}(\hat t) \ 
C_{i2}(\hat t) 
\Big)
\binom{\av{D^+|\bar h_{v'} \sla{n} P_L h_v|\bar B^0}}
{\av{D^+|\bar h_{v'} \sla{n} P_R h_v|\bar B^0}}
\av{L^-| \bar \chi_n^{(d)}(t\bar n)\frac{\sla{\bar n}}{2}  P_L \chi_n^{(u)}(0) | 0} \, . \label{eq:MEQCDSCET1}
}
The individual factors in this equation can be expressed in terms of non-perturbative
objects. For the term in the middle one has, in the notation of~\REF{Huber:2016xod},
\eq{
\binom{\av{D^+|\bar h_{v'} \sla{n} P_L h_v|\bar B^0}}
{\av{D^+|\bar h_{v'} \sla{n} P_R h_v|\bar B^0}} =
\left(
\begin{array}{cc}
C_{FF}^D & C_{FF}^{ND} \\
C_{FF}^{ND} & C_{FF}^D 
\end{array}
\right)^{-1}
\binom{\av{D^+|\bar c \sla{n} P_L b|\bar B^0}}
{\av{D^+|\bar c \sla{n} P_R b|\bar B^0}} \, , \label{eq:matchingheavy}
}
while for the last term we get
\eqa{
\av{L^-(k)| \bar \chi_n^{(d)}(t\bar n)\frac{\sla{\bar n}}{2} P_L \chi_n^{(u)}(0) | 0} & = &
C_{q\bar q}^{-1}
\av{L^-(k)| \bar d_n(t\bar n)\frac{\sla{\bar n}}{2} P_L u_n(0) | 0} \nonumber\\[0.2em]
& \equiv & C_{q\bar q}^{-1}\, \hat\Phi_L(k,t) = C_{q\bar q}^{-1}\, k^- \int_0^1 du\ e^{i u t k^-} \hat\Phi_L(k,u) \, , \label{eq:matchinglight}
}
where for the dimeson state we let the variable $k$ represent the pair $(k_1,k_2)$. 
The matching coefficient $C_{q\bar q}$ accounts for the difference between the QCD and SCET current as discussed in~\REF{Beneke:2009ek}, and we have introduced a Fourier transformation from the variable $\hat t = t \, k^-$ to the momentum fraction $u \in [0,1]$,
\eq{
\hat\Phi_L(k,t) \equiv k^- \int_0^1 du\ e^{i u t k^-} \hat\Phi_L(k,u)\, .
}
Substituting~(\ref{eq:matchingheavy}) and~(\ref{eq:matchinglight}) into~(\ref{eq:MEQCDSCET1}) we obtain
\eqa{
\av{Q_i} \! \! &=& \! \! k^- \! \int_0^1 \! du \! \int \! d\hat t \, e^{i u t k^-} \Big( C_{i1}(\hat t) \ C_{i2}(\hat t) \Big)
\left(
\begin{array}{cc}
C_{FF}^D & C_{FF}^{ND} \\
C_{FF}^{ND} & C_{FF}^D 
\end{array}
\right)^{-1} \!
\binom{\av{D^+|\bar c \sla{n} P_L b|\bar B^0}}
{\av{D^+|\bar c \sla{n} P_R b|\bar B^0}}
C_{q\bar q}^{-1} \hat\Phi_L(k,u) . \nonumber\\[0.2em]
&&\label{eq:MEQCDSCET2}
}
The Fourier transform of the matching coefficients $C_{ik}(\hat t)$ can be identified with the quantities $H_{i1}^{(\prime)}(u)$ defined in~\REF{Huber:2016xod},
\eq{
\Big( H_{i1}(u) \ H_{i1}^{\prime}(u) \Big) = \int d\hat t \, e^{i u t k^-} \Big( C_{i1}(\hat t) \ C_{i2}(\hat t) \Big) \, .
}
The entire calculation in~\REF{Huber:2016xod} was performed in momentum space and therefore the $H_{i1}^{(\prime)}$ (and $\hat T_i^{(\prime)}$ below) were directly obtained as a function of $u$, see also the discussion in~\REF{Beneke:2008ei}.
Together with the form-factor relation
\eq{
{\av{D^+|\bar c \sla{n} P_L b|\bar B^0}} =
{\av{D^+|\bar c \sla{n} P_R b|\bar B^0}} =
F_n^{B\to D}(k^2)\equiv F_n^{B\to D} \, ,
\label{eq:Fn}
}
\Eq{eq:MEQCDSCET2} becomes
\eqa{
\av{Q_i} &=&  k^- \, F_n^{B\to D} \, \int_0^1 \! du \Big( H_{i1}(u) \ H_{i1}^{\prime}(u) \Big)
\left(
\begin{array}{cc}
C_{FF}^D & C_{FF}^{ND} \\
C_{FF}^{ND} & C_{FF}^D 
\end{array}
\right)^{-1} \!
\binom{1}{1}
\, C_{q\bar q}^{-1} \hat\Phi_L(k,u) . \label{eq:MEQCDSCET3}
}
Following again~\REF{Huber:2016xod} we see that~(\ref{eq:MEQCDSCET3}) contains the hard functions
$\hat T_i^{(\prime)}(u)$ which are related to the $H_{i1}^{(\prime)}(u)$ via
\eq{
\Big( \hat T_{i}(u) \ \hat T_{i}^{\prime}(u) \Big) = \Big( H_{i1}(u) \ H_{i1}^{\prime}(u) \Big)
\left(
\begin{array}{cc}
C_{FF}^D & C_{FF}^{ND} \\
C_{FF}^{ND} & C_{FF}^D 
\end{array}
\right)^{-1}
\, C_{q\bar q}^{-1} \, .
}
Therefore,
\eqa{
\av{Q_i} &=&  k^- \, F_n^{B\to D} \, \int_0^1 \! du \Big( \hat T_{i}(u) \ \hat T_{i}^{\prime}(u) \Big)
\binom{1}{1} \, \hat\Phi_L(k,u) \, . \label{eq:MEQCDSCET4}
}
In other words, the hard functions in the factorization formula are universal for all $\bar B^0 \to D^+ L^-$
decays. As expected, they enter the $\bar B^0 \to D^+L^-$ transition only
in the combination $T_i(u) = \hat T_i(u) + \hat T^{\prime}_i(u)$ and we arrive at the factorization formula for the matrix element:
\eq{
\av{Q_i} = k^- \, F_n^{B\to D} \int_0^1 du\ T_i(u)\ \hat\Phi_L(k,u) \ . \label{eq:factorization}
}
At order $\alpha_s^0$, the hard functions are given by
\eq{
T_1(u) = \cO (\alpha_s)\ ,\quad T_2(u) = 1 + \cO (\alpha_s)\ ,
}
and, in particular, the matrix elements depend only on the local limit of the LCDAs, i.e.\ decay constants for $L=meson$ or timelike form factors for $L=dimeson$.
The functions $T_i(u)$ have been computed up to $\cO (\alpha_s^2)$ in \REF{Huber:2016xod}, and we will use these results to produce NNLO predictions for the amplitudes.

\subsection{Form factors and generalized LCDAs}

The form factor $F_n^{B\to D}$ as defined in~\Eq{eq:Fn} can be easily related to the traditional form factors $F_{0,+}^{B\to D}$ (e.g.~\cite{Beneke:2000wa}):
\eq{
{\av{D^+(q)|\bar c \gamma^\mu b|\bar B^0(p)}} = F_+(k^2) \bigg[ (p+q)^\mu - \frac{m_B^2-m_D^2}{k^2} k^\mu\bigg] + F_0(k^2)\, \frac{m_B^2-m_D^2}{k^2} k^\mu\ .
}
Contracting this matrix element with $n_\mu/2$, one finds
\eq{
k^- F_n^{B\to D} = \frac{m_B^2-m_D^2}{2} \,\big[F_+(k^2)+F_0(k^2)\big] + \cO(k^2/m_B^2)
= (m_B^2-m_D^2)\,F_0(k^2) + \cO(k^2/m_B^2)\ ,
}
where the second equality results from the kinematic constraint $F_+(0)=F_0(0)$.

The LCDAs $\hat \Phi_L(k,u)$ have been defined in~\Eq{eq:matchinglight}, but for reasons that will become clear below we choose to express them in terms of ``unhatted'' LCDAs in the following way:
\eqa{
\hat\Phi_P(k,u) &=& \frac{if_P}{4}\, \Phi_P(u)\ ,\\
\hat\Phi_V(k,u) &=& \frac{f_V}{4}\, \Phi_V(u)\ , \\
\hat\Phi_{\pi\pi}(k_1,k_2,u) &=& -\frac{1}{2\sqrt2}\, \Phi_{\pi\pi}(u,k^2,\theta_\pi)\ ,
\label{eq:PhiHatpipi}\\
\hat\Phi_{K\pi}(k_1,k_2,u) &=& -\frac1{2\sqrt2}
\Phi_{K\pi}(u,k^2,\theta_\pi) \ ,
\label{eq:PhiHatKpi}
} 
where $P=\{\pi,K\}$ and $V=\{\rho,K^*\}$.
The decay constants $f_{P,V}$ are defined by:
\eq{
\av{P(k)|\bar x \gamma_\mu \gamma_5 u |0} = -i f_P k_\mu\ ,
\quad
\av{V(k,\varepsilon)|\bar x \gamma_\mu u |0} = f_V m_V \varepsilon^*_\mu\ ,
}
and thus,\footnote{For vector mesons and vector resonances $R$, we use the fact that $\varepsilon_\| \cdot \bar n = k^-/m_R$ and $\varepsilon_t \cdot \bar n = \varepsilon_0 \cdot \bar n = \varepsilon_\perp \cdot \bar n =0$. Thus the decay constant appearing really is $f_V=f^\|_V$.}
\eq{
\av{P(k)|\bar x \,\sla{\bar n}\, \gamma_5 u |0} = -i f_P k^-\ ,
\quad
\av{V(k,\varepsilon_\|)|\bar x \,\sla{\bar n}\, u |0} = f_V^\| \,k^- \ .
}
For vector mesons we have adopted the phase convention of~Refs.~\cite{Cheng:2017smj,Descotes-Genon:2019bud}, which differs from~\cite{Beneke:2000wa} but is more convenient when regarding the vector meson as a resonance.
The light-meson LCDAs $\Phi_P(u)$ and $\Phi_V(u)$ defined in this way
coincide with the ones used in~Refs.~\cite{Beneke:2000wa,Beneke:2000ry}\,\footnote{In \cite{Beneke:2000wa} the notation $\phi_p(u)$ and $\phi_\|(u)$ is used, while $\Phi_V=\Phi_\|$ in~\cite{Beneke:2000ry}. See App.\,A of\cite{Beneke:2000wa} and Sec.\,2.3.2 of~\cite{Beneke:2000ry}.}, satisfying the normalization
\eq{
\int_0^1 du\, \Phi_P(u) = \int_0^1 du\, \Phi_V(u) = 1\ .
}
Concerning the dimeson LCDAs, their local limit is given by the pion and $K\pi$ timelike form factors:
\eqa{
\av{\pi^-(k_1)\pi^0(k_2)| \bar d \gamma_\mu u | 0} &=& -\sqrt2\, F_\pi(k^2)\, \overline k_\mu\ ,
\label{eq:Fpi}\\[2mm]
\av{K^-(k_1)\pi^0(k_2)| \bar s \gamma_\mu u | 0} &=& -\frac{f^{K\pi}_+(k^2)}{\sqrt2}\,\overline k_\mu - \frac{\Delta m_{K\pi}^2}{\sqrt2 k^2}\,f_0^{K\pi}(k^2)\,k_\mu\ ,
\label{eq:FKpi}
}
where $F_\pi(k^2)$ denotes the electromagnetic form factor with $F_\pi(0)=1$, and $f_{0,+}^{K\pi}(k^2)$ denote the scalar and vector $\bar K^0\pi^-$ form factors, respectively. Thus the various signs and factors of $\sqrt2$ are isospin coefficients. These form factors in the timelike region can be extracted directly from data (see e.g.~\cite{Fujikawa:2008ma,Cheng:2017smj,Epifanov:2007rf,Descotes-Genon:2019bud}).

These local limits determine that the ``unhatted'' LCDAs in the right-hand side of Eqs.~(\ref{eq:PhiHatpipi}) and~(\ref{eq:PhiHatKpi}) are normalized to:
\eqa{
\int_0^1 du\, \Phi_{\pi\pi}(u,k^2,\theta_\pi) &=&
\cos\theta_\pi\,\beta_\pi(k^2) F_\pi(k^2)\ ,
\label{eq:normPhipipi}\\
\int_0^1 du\, \Phi_{K\pi}(u,k^2,\theta_\pi) &=& 
\cos\theta_\pi \frac{\sqrt{\lambda_{K\pi}(k^2)}}{2k^2} f_+^{K\pi}(k^2) + \frac{\Delta m_{K\pi}^2}{2k^2} f_0^{K\pi}(k^2)\ ,
\label{eq:normPhiKpi}
}
where $\beta_\pi(k^2) \equiv \sqrt{1-4m_\pi^2/k^2}$.
One can see that in the $\pi\pi$ case, only a $P$-wave component arises, while in the $K\pi$ case there is also an $S$-wave component proportional to the mass difference $\Delta m_{K\pi}^2 = m_K^2-m_\pi^2$. We also see that in the limit $m_K\to m_\pi$, where $\sqrt{\lambda_{K\pi}(k^2)}\to k^2 \beta_\pi(k^2)$, the $K\pi$ case reduces to that of $\pi\pi$ up to isospin factors.
The definition of the dipion LCDA $\Phi_{\pi\pi}$ used here thus agrees with the ones used in~Refs.~\cite{Polyakov:1998ze,Krankl:2015fha,Cheng:2017sfk}. The $K\pi$ LCDA $\Phi_{K\pi}$, however, has not been discussed in the literature in any depth.

The LCDAs can be expanded in Gegenbauer coefficients:
\eq{
\Phi_L(u) = 6 u\bar u\,
\sum_{n=0}^\infty\alpha_n^L\,C_n^{3/2}(u-\bar u)\ ,
\label{eq:PhiL}
} 
where $C_0^{3/2}(x)=1$, $C_1^{3/2}(x)=3x$, $C_2^{3/2}(x)=3(5x^2-1)/2$, etc.
The normalization of the LCDAs determines $\alpha_0^L = \int_0^1 du\,\Phi_L(u)$, such that $\alpha_0^P = \alpha_{0}^V = 1$ and $\alpha_0^{M\pi}$ are given by~Eqs.~(\ref{eq:normPhipipi}) and~(\ref{eq:normPhiKpi}). 
For $L=\{\rho,\pi\pi\}$, $\alpha^L_{n\text{\,odd}}=0$ due to C-parity in the isospin limit.
In the dimeson case the Gegenbauer coefficients depend on $(k^2,\theta_\pi)$, and can be expanded in partial waves. For $L=\pi\pi$,
\eq{
\alpha_n^{\pi\pi}(k^2,\theta_\pi) = \sum_{\ell=1,3,\cdots}^{n+1}
B^{\pi\pi}_{n\ell}(k^2)\, P_\ell(\cos\theta_\pi) 
\qquad (n\ \text{even})\ ,
\label{eq:alphapipi}
}
where the normalization fixes
\eq{
B^{\pi\pi}_{01}(k^2) = \beta_\pi(k^2) F_\pi(k^2)\ .
\label{eq:B01pipi}
}
This definition for the coefficients $B_{n\ell}^{\pi\pi}$ differs from~Refs.~\cite{Cheng:2017sfk,Polyakov:1998ze}\,\footnote{In these references the notation $B_{n\ell}^{\pi\pi} = B_{n\ell}^{\|}$ is used.} by the factor $\beta_\pi(k^2)$. We choose this definition in order to have a homogeneous notation between the $\pi\pi$ and $K\pi$ cases.
For $L=K\pi$, we write:
\eq{
\alpha_n^{K\pi}(k^2,\theta_\pi) = \sum_{\ell=0}^{n+1}
B^{K\pi}_{n\ell}(k^2)\, P_\ell(\cos\theta_\pi)  \qquad (\text{all}\ n)\ .
\label{eq:alphaKpi}
}
In this case the normalization fixes
\eq{
B^{K\pi}_{00}(k^2) = \frac{\Delta m_{K\pi}^2}{2 k^2} f_0^{K\pi}(k^2)\ ,
\qquad
B^{K\pi}_{01}(k^2) = \frac{\sqrt{\lambda_{K\pi}(k^2)}}{2k^2}f_+^{K\pi}(k^2)\ .
\label{eq:B00B01Kpi}
}
We will only consider Gegenbauer expansions up to $n=2$, as in~\REF{Huber:2016xod}.

\subsection{Factorized $\bar B\to D^+L^-$ amplitudes}

We can now write down the amplitudes
\eq{
\A(\bar B\to D^+L^-) = 
\frac{4 G_F}{\sqrt2} \,V_{ux}^* V_{cb}\ k^-\,F_n^{B\to D}
\int_0^1 du\ \big( C_1 T_1(u) + C_2 T_2(u) \big) \hat\Phi_L(k,u)\ .
}
With the previous considerations, we have:
\eqa{
\A(\bar B\to D^+P^-)
&=& 
i\,\frac{G_F}{\sqrt2} \,V_{ux}^* V_{cb}\ (m_B^2-m_D^2)\ F_0^{B\to D}(m_P^2)  f_P\ a_1(D^+P^-)
\ , 
\label{eq:A(B->DP)}\\
\A(\bar B\to D^+V^-)
&=& 
\frac{G_F}{\sqrt2} \,V_{ux}^* V_{cb}\ (m_B^2-m_D^2)\ F_0^{B\to D}(m_V^2)  f^\|_V\ a_1(D^+V^-)
\ , 
\label{eq:A(B->DV)}\\
\A(\bar B\to D^+M^-\pi^0)
&=& 
-\,G_F \,V_{ux}^* V_{cb}\ (m_B^2-m_D^2)\ F_0^{B\to D}(k^2) \ a_1(D^+M^-\pi^0)
\ .
\label{eq:A(B->DMpi)}
}
The coefficients $a_1(D^+L^-)$ correspond to the same coefficients as in Refs.~\cite{Beneke:2000ry,Huber:2016xod} for the cases $L=P,V$, which we generalize here,
\eq{
a_1(D^+L^-) = \int_0^1 du\ \big( C_1 T_1(u) + C_2 T_2(u) \big) \,\Phi_L(u) \ ,
\label{eq:a1}
}
noting that in the cases $L=M\pi$ the dipion LCDAs $\Phi_L(u,k^2,\theta_\pi)$ depend on the two kinematic variables $k_1$ and $k_2$. The amplitudes for $B\to DP$ and $B\to DV$ in Eqs.~(\ref{eq:A(B->DP)}) and~(\ref{eq:A(B->DV)}) agree with the literature\cite{Beneke:2000ry,Huber:2016xod}, after accounting for the phase redefinition in the vector-meson state.

\bigskip

After expanding the LCDAs in Gegenbauer coefficients, only the convolutions of the hard coefficients $T_i(u)$ with the Gegenbauer polynomials are needed. 
We denote these by
\eq{
\int_0^1 du\, T_i(u,\mu)\, 6 u\bar u\, C_n^{3/2}(u-\bar u) = \V_{in}(\mu)\ .
}
In the notation of~\REF{Huber:2016xod}, with $z_c = m_c^2/m_b^2$ and with the masses in the pole scheme,
\eq{
\V_{in}(\mu) = \sum_{m\ge 0} \bigg( \frac{\alpha_s(\mu)}{4\pi} \bigg)^m
\Big[ V_{in}^{(m)}(\mu) + V_{in}^{\prime(m)}(\mu) \sqrt{z_c} \Big] \ ,
}
where
\eqa{
&& V_{10}^{(0)}(\mu)=V_{10}^{\prime(0)}(\mu)
=V_{20}^{\prime(0)}(\mu)=0\ ,
\quad V_{20}^{(0)}(\mu)=1\ , \\[2mm]
&& V_{in}^{(0)}(\mu) = V_{in}^{\prime(0)}(\mu)=0
\quad \text{for} \quad n\ge 1\ .
}
With these definitions, the amplitudes $a_1(D^+L^-)$ in~\Eq{eq:a1} are written as
\eq{
a_1(D^+L^-) =
\sum_{n\ge 0} \alpha_n^L\,
\big[C_1(\mu) \V_{1n}(\mu)+C_2(\mu)\V_{2n}(\mu)\big]
\equiv \sum_{n\ge 0} \alpha_n^L\, {\cal G}_n(\mu)
\ .
\label{eq:a1NNLO}
}
Inserting this form of the amplitudes $a_1$ into~\Eq{eq:A(B->DMpi)} gives the $B\to D M\pi$ amplitudes in terms of the coefficients $\alpha_n^{M\pi}(k^2,\theta_\pi)$.

Finally, adopting the partial-wave decompositions for the Gegenbauer coefficients given in Eqs.~(\ref{eq:alphapipi}) and~(\ref{eq:alphaKpi}), one obtains the partial-wave amplitudes from~\Eq{eq:APW}, 
\eq{
\A^{(\ell)}(k^2) = -\,G_F \,V_{ux}^* V_{cb}\ (m_B^2-m_D^2)\ F_0^{B\to D}
\sum_{n\ge{\rm{max}}(\ell-1,0)} B_{n\ell}^L(k^2)\, {\cal G}_n(\mu)\ .
\label{eq:A(B->DMpi)PWE}
}
With the coefficients $\V_{in}$ known at two loops, Eqs.~(\ref{eq:A(B->DMpi)}) [together with~ (\ref{eq:a1NNLO})] and~(\ref{eq:A(B->DMpi)PWE}) [with the normalizations of $B_{0\ell}^{M\pi}(k^2)$ from~Eqs.~(\ref{eq:B01pipi}) and (\ref{eq:B00B01Kpi})] give the leading-power amplitude for the $B\to D M\pi$ decay at NNLO in QCD, in the region of small~$k^2$.  This constitutes one of the main results of the present paper.

\subsection{Numerical size of NLO and NNLO terms}
\label{sec:numerics}

Perturbative corrections to the hard functions in the leading-power amplitude are contained in the quantities
\eq{
{\cal G}_n(\mu) \equiv \big[C_1(\mu) \V_{1n}(\mu)+C_2(\mu)\V_{2n}(\mu)\big]
\label{eq:CiVi}
}
in e.g.~\Eq{eq:a1NNLO}, with $n$ referring to the Gegenbauer expansion. These quantities contain all the short-distance information available from the decay amplitude. At tree-level, there are no contributions for $n>0$, and thus the information from higher Gegenbauer moments enters through $\alpha_s(m_b)$ corrections. In order to gauge the importance of higher Gegenbauer contributions and thus to establish how sensitive these amplitudes are to the hadronic structure of the dimeson systems beyond the asymptotic limit, one needs to pay attention to perturbative effects. Hence here we recollect briefly the numerical size of the quantities in~\Eq{eq:CiVi}.
 
In the calculation of the Wilson coefficients $C_i(\mu)$ we take the two-loop SM matching conditions and running from~\REF{Bobeth:1999mk} and set the electroweak matching scale to $\mu_0=160\GeV$. We then set $\mu_b=m_{b,pole}=4.78\GeV$ for the SCET matching scale, and $m_{c,pole}=1.67\GeV$. We then have
\eqa{
C_1(\mu_b) &=&
-0.59 + 16.8\ \widetilde\alpha_s + 145.4\ \widetilde\alpha_s^2
= -0.26\ ,\\[1mm]
C_2(\mu_b) &=&
1.03 -  1.70 \ \widetilde\alpha_s  + 9.06\ \widetilde\alpha_s^2
= 1.01\ ,
}
where
\eq{
\widetilde\alpha_s  \equiv  \alpha_s(\mu_b)/(4\pi) = 0.01726\ .
}
The QCDF coefficients $\V_{in}$ up to NNLO are taken from~\REF{Huber:2016xod}. We find,
\eqa{
\V_{10} &=& 
(-5.36 - i\,1.91)\ \widetilde\alpha_s + 
(-59.41 - i\,80.13)\ \widetilde\alpha_s^2 
= -0.1103 - i\,0.0569\\[1mm]
\V_{11} &=& 
(1.25 - i\,2.93)\ \widetilde\alpha_s + 
(182.63 - i\,87.94)\ \widetilde\alpha_s^2 
= 0.076 - i\,0.077\\[1mm]
\V_{12} &=& 
(-0.22 + i\,0.16)\ \widetilde\alpha_s + 
(-28.99 - i\,2.48)\ \widetilde\alpha_s^2 
= -0.0124 + i\,0.0021\\[1mm]
\V_{20} &=& 1 +  
(87.85 + i\,73.93)\ \widetilde\alpha_s^2 
= 1.0262 + i\,0.0220\\[1mm]
\V_{21} &=&  
(-59.58 + i\,56.62)\ \widetilde\alpha_s^2 
= -0.018 + i\,0.017\\[1mm]
\V_{22} &=&  
(-7.49 - i\,21.57)\ \widetilde\alpha_s^2 
= -0.0022 - i\,0.0064
}
Combining the Wilson coefficients with the QCDF coefficients order by order in $\alpha_s$, the relevant short-distance quantities in the QCDF amplitude are given by 
\eqa{
{\cal G}_0(\mu_b) &=&
1.034 + (1.488 + i\,1.134) \ \widetilde\alpha_s
+ (45.15 + i\,91.96) \ \widetilde\alpha_s^2
\nonumber \\
&=& 1.034_{\rm LO} + (0.026 + i\,0.020)_{\rm NLO} + (0.013 + i\,0.027)_{\rm NNLO}
\nonumber \\
&=& 1.07 + i\,0.047\ , \\[3mm]
{\cal G}_1(\mu_b)   &=&
(-0.74 + i\,1.74) \ \widetilde\alpha_s
+ (-149.1 + i\,61.5) \ \widetilde\alpha_s^2
\nonumber \\
&=& (-0.013 + i\,0.030)_{\rm NLO} +  (-0.044 + i\,0.018)_{\rm NNLO} \nonumber \\
&=& -0.057 + i\,0.048\ , \\[3mm]
{\cal G}_2(\mu_b)   &=&
(0.132 - i\,0.096) \ \widetilde\alpha_s
+ (5.73 - i\,18.12) \ \widetilde\alpha_s^2
\nonumber \\
&=& (0.0023 - i\,0.0017)_{\rm NLO} +  (0.0017 - i\,0.0054)_{\rm NNLO} \nonumber \\
&=& 0.0040 - i\,0.0071 \ .
}
One can see that, while NLO corrections are typically tiny (around $3\%$) and characteristic of a tree-level decay amplitude, NNLO corrections are large relative to the NLO. The reason for this behaviour is the vanishing colour factor of $\V_{2n}$ at order ${\cal O}(\alpha_s)$, leaving only the contribution of the smaller Wilson coefficient $C_1$ at this order. Thus, when considering contributions from the higher Gegenbauer moments (which are genuinely NLO), NNLO contributions happen to be very important. This is especially relevant for the $DK \pi$ case, which receives contributions from $\alpha_1$.

Substituting these results into the amplitude $a_1$, we find (up to $n=2$):
\eqa{
a_1(D^+L^-) &=&
\big[1.034\,\alpha_0^L\big]_{\rm LO}\nonumber\\
&& + \big[(0.026 + i\,0.020)\alpha_0^L + (-0.013 + i\,0.030)\alpha_1^L + (0.0023 - i\,0.0017)\alpha_2^L\big]_{\rm NLO}
\nonumber\\
&& + \big[(0.013 + i\,0.027)\alpha_0^L + (-0.044 + i\,0.018)\alpha_1^L + (0.0017 - i\,0.0054)\alpha_2^L\big]_{\rm NNLO}
\nonumber\\[2mm]
&=& (1.073 + i\, 0.047)\alpha_0^L + (-0.057 + i\,0.048)\alpha_1^L + (0.0040 - i\,0.0071)\alpha_2^L\ .
}
For the squared amplitude appearing in the differential rate we have,
normalizing to the leading Gegenbauer contribution and defining $\hat\alpha_i^L\equiv \alpha_i^L/\alpha_0^L$,
\eqa{
|a_1(D^+L^-)|^2 &=& |\alpha_0^L|^2 \,\Big\{
1.07_{\rm LO}\nonumber\\
&& + \big[0.053
- 0.026 \,\re\,\hat\alpha_1^L - 0.062\, \im\,\hat\alpha_1^L
+ 0.0047 \,\re\,\hat\alpha_2^L + 0.0034\, \im\,\hat\alpha_2^L
\big]_{\rm NLO}
\nonumber\\
&& + \big[0.029
- 0.091 \,\re\,\hat\alpha_1^L - 0.040\, \im\,\hat\alpha_1^L
+ 0.0036 \,\re\,\hat\alpha_2^L + 0.011\, \im\,\hat\alpha_2^L
\big]_{\rm NNLO}
\Big\}
\nonumber\\[2mm]
&=& 1.15 |\alpha_0^L|^2\,\Big\{
1
- 0.10\,\re\,\hat\alpha_1^L - 0.09\,\im\,\hat\alpha_1^L
+ 0.007\,\re\,\hat\alpha_2^L + 0.014\,\im\,\hat\alpha_2^L
\Big\}\ .
}
One can see that $n=1$ corrections are of the order of $10\%$ compared to the leading $n=0$ terms, while $n=2$ corrections are one order of magnitude smaller. In each case, NNLO corrections are essential, to the point that the $\re\,\hat\alpha_1^L$ term is dominated by the NNLO contribution.
Since for the $D\pi\pi$ channel $\hat \alpha_1^{\pi\pi}=0$, this decay should be well approximated by the asymptotic term:
\eq{
|a_1(D^+\pi^-\pi^0)|^2 \simeq 1.15 |\alpha_0^{\pi\pi}|^2\ ,
}
up to corrections at the $1\%$ level, and with perturbative NLO and NNLO corrections accounting for $\sim 7\%$ of the numerical coefficient $1.15$. 
However, for the $DK\pi$ channel, corrections from $\alpha_1$ are at the level of $10\%$ (depending on the size of the nonperturbative Gegenbauer coefficients $\hat\alpha_1$), and possibly more important than perturbative corrections to the leading asymptotic contribution from $\alpha_0$.
For this channel, we will write,
\eq{
|a_1(D^+K^-\pi^0)|^2 \simeq 1.15  (1 + \xi_{K\pi}) |\alpha_0^{K\pi}|^2\ ,
}
with
\eq{
\xi_{K\pi} \equiv - 0.10\,\re\,\hat\alpha_1^{K\pi} - 0.09\,\im\,\hat\alpha_1^{K\pi}
+ 0.007\,\re\,\hat\alpha_2^{K\pi} + 0.014\,\im\,\hat\alpha_2^{K\pi} \sim \cO(0.1)\ .
}
These simplified expressions might be useful to understand the importance of higher-order effects in observables.

\subsection{Modeling the dimeson system}

The coefficients $B_{n\ell}^L(k^2)$ determine the $k^2$ spectrum of each partial-wave amplitude, and can eventually be extracted from data. For example, the $n=0$ coefficients $B_{01}^{\pi\pi}(k^2)$, $B_{00}^{K\pi}(k^2)$ and $B_{01}^{K\pi}(k^2)$ are given by the timelike form factors in Eqs.~(\ref{eq:Fpi}) and~(\ref{eq:FKpi}), which can be extracted from $\tau$-decay spectra~\cite{Fujikawa:2008ma,Cheng:2017smj,Epifanov:2007rf,Descotes-Genon:2019bud}. Still, modeling these functions is necessary and useful. For instance, data extractions are essentially fits to the parameters of a model, and a model that fits well the data then becomes a proxy for the data itself. 

The most common models are those where it is assumed that the dimeson is produced through a coupling to a resonance created via the weak current. We will use a class of such models, in particular the one employed in~Refs.~\cite{Cheng:2017smj,Descotes-Genon:2019bud}, which contains the models used in the experimental analyses of ~Refs.~\cite{Fujikawa:2008ma,Epifanov:2007rf}, and which describe the form factor data very well. In the case of $\pi\pi$ the data is well described with inclusion of only $J^P=1^-$ resonances ($\rho$, $\rho'$, $\rho''$)~\cite{Fujikawa:2008ma,Cheng:2017smj}, while in the case of $K\pi$, both $J^P=1^-$ ($K^*(892)$, $K^*(1410)$) and $J^P=0^+$ ($K_0^*(800)$, $K_0^*(1430)$) may be required~\cite{Epifanov:2007rf,Descotes-Genon:2019bud}.

Thus it seems to suffice to consider $\ell=\{0,1\}$ only. Following the same derivation as in~\REF{Descotes-Genon:2019bud}, we find:
\eqa{
B_{n0}^{M\pi}(s) &=& \sum_{R_0} \frac{m_{R_0}\, f_{R_0}\, g_{R_0M\pi}\, e^{i\varphi_{R_0}}}{\sqrt{2} [m_{R_0}^2 - s- i\,\sqrt{s}\, \Gamma_{R_0}(s)]}\, \alpha_n^{R_0}\ ,
\label{eq:Bn0}
\\
B_{n1}^{M\pi}(s) &=& \frac{\sqrt{\lambda_{M\pi}(s)}}{s} \sum_{R} \frac{m_{R}\, f_{R}\, g_{RM\pi}\, e^{i\varphi_{R}}}{\sqrt{2} [m_{R}^2 - s- i\,\sqrt{s}\,\Gamma_{R}(s)]}\, \alpha_n^{R}\ .
\label{eq:Bn1}
}
The sums run over $R_0=0^+$ and $R=1^-$ resonances. The decay constants $f_R$ correspond to that of vector mesons defined above, while $f_{R_0}$ are defined analogously:
\eq{
\av{R_0(k)|\bar x \gamma_\mu u |0} = f_{R_0}\, k_\mu\ .
}
The quantities $\alpha_n^{R,R_0}$ correspond (as the notation suggests) to the Gegenbauer moments of the LCDAs of the resonances. The strong couplings $g_{RM\pi}$ and $g_{R_0M\pi}$ are defined by
\eq{
\av{M^-\pi^0|R(k,\varepsilon)} = g_{RM\pi}\, e^{i\varphi_R} \,\bar k\cdot \varepsilon\ \ ;
\qquad
\av{M^-\pi^0|R_0} = m_{R_0}\, g_{R_0M\pi}\, e^{i\varphi_{R_0}}\ ,
}
and determine the partial widths of the resonances:
\eq{
\Gamma_R^{\rm tot} = \frac{g_{RM\pi}^2}{48\pi} \frac{\lambda_{M\pi}^{3/2}(m_R^2)}{m_R^5} \frac1{\B(R\to M^-\pi^0)} \ \ ;\quad
\Gamma_{R_0}^{\rm tot} = \frac{g_{R_0M\pi}^2}{48\pi} \frac{\lambda_{M\pi}^{1/2}(m_{R_0}^2)}{m_{R_0}^3} \frac1{\B(R_0\to M^-\pi^0)}\ .
}
Finally, the $s$-dependent widths are given by
\eq{
\Gamma_{R}(s) = \Gamma_R^{\rm tot} \bigg[ \frac{\lambda_{M\pi}(s)}{\lambda_{M\pi}(m_R^2)} \bigg]^{3/2}
\frac{m_R^5}{s^{5/2}} \theta(s-s_{\rm th}) \ \ ;\quad
\Gamma_{R_0}(s) = \Gamma_{R_0}^{\rm tot} \bigg[ \frac{\lambda_{M\pi}(s)}{\lambda_{M\pi}(m_{R_0}^2)} \bigg]^{1/2}
\frac{m_{R_0}^3}{s^{3/2}} \theta(s-s_{\rm th}) \ .
}
It will also be useful to write the strong couplings in terms of the $s$-dependent widths,
\eq{
g_{RM\pi}^2 = \frac{48\pi s^{5/2} \Gamma_R(s)}{\lambda_{M\pi}^{3/2}(s)} \B(R\to M^-\pi^0) \ ,
\quad
g_{R_0M\pi}^2 = \frac{48\pi s^{3/2} \Gamma_{R_0}(s)}{\lambda_{M\pi}^{1/2}(s)} \B(R_0\to M^-\pi^0) \ .
\label{eq:gRMpi}
}
Thus, in these models the amplitudes are determined by a few parameters: the masses and partial widths of the resonances (taken, e.g. from the PDG~\cite{Tanabashi:2018oca}), and the Gegenbauer moments $\alpha_n^{R,R_0}$, which are to be regarded as the free parameters of the model. As will be shown below, these will correspond, in the narrow width limit, to the Gegenbauer moments of stable vector and scalar mesons.
Since the normalization of LCDAs for stable mesons has been chosen such that $\alpha_0^M=1$ (see text below~\Eq{eq:PhiL}), with this identification we have from~Eqs.~(\ref{eq:B01pipi}), (\ref{eq:B00B01Kpi}) and~(\ref{eq:Bn1}) that:
\eqa{
B_{01}^{\pi\pi}(s) &=& \beta_\pi(s) F_\pi(s) = \beta_\pi(s) \sum_{R} \frac{m_{R}\, f_{R}\, g_{R\pi\pi}\, e^{i\varphi_{R}}}{\sqrt{2} [m_{R}^2 - s- i\,\sqrt{s}\,\Gamma_{R}(s)]}\ ,
\label{eq:B01pipiModel}\\
B_{01}^{K\pi}(s) &=& \frac{\sqrt{\lambda_{K\pi}(s)}}{2s}f_+^{K\pi}(s)
=\frac{\sqrt{\lambda_{K\pi}(s)}}{s} \sum_{R} \frac{m_{R}\, f_{R}\, g_{RK\pi}\, e^{i\varphi_{R}}}{\sqrt{2} [m_{R}^2 - s- i\,\sqrt{s}\,\Gamma_{R}(s)]}\label{eq:B01piKModel}\ ,
}
which agrees with the model expressions for $F_\pi(s)$ and $f_+(s)$ in~Refs.~\cite{Cheng:2017smj,Descotes-Genon:2019bud}\,\footnote{
The notation obscures somewhat the agreement with $f_+$ in~\REF{Descotes-Genon:2019bud}. Note that:
$$f_+^{K^-\pi^+}(s)=-f_+^{\bar K^0\pi^-}(s)\equiv - f_+^{K\pi}(s)
= -\sum_{R} \frac{\sqrt{2}\,m_{R}\, f_{R}\, g_{RK\pi}\, e^{i\varphi_{R}}}{[m_{R}^2 - s-i\,\sqrt{s}\,\Gamma_{R}(s)]}
= -\sum_{R} \frac{m_{R}\, f_{R}\, g_{RK^-\pi^+}\, e^{i\varphi_{R}}}{[m_{R}^2 - s- i\,\sqrt{s}\,\Gamma_{R}(s)]}\ , $$
where $g_{RK^-\pi^+} = \sqrt{2}\,g_{RK\pi}$ is the strong coupling of the $R\to K^-\pi^+$ decay appearing in~\REF{Descotes-Genon:2019bud}. 
}.
In the case of the $S$-wave,
\eq{
B^{K\pi}_{00}(s) = \frac{\Delta m_{K\pi}^2}{2 k^2} f_0^{K\pi}(s) = 
\sum_{R_0} \frac{m_{R_0}\, f_{R_0}\, g_{R_0K\pi}\, e^{i\varphi_{R_0}}}{\sqrt{2} [m_{R_0}^2 - s- i\,\sqrt{s}\, \Gamma_{R_0}(s)]}\ ,
}
which agrees with~Refs.~\cite{Epifanov:2007rf,Descotes-Genon:2019bud} when
\eq{
f_+(0)\, \xi_{R_0} = -\frac{\sqrt{2}\,m_{R_0}\, f_{R_0}\, g_{R_0K\pi}\, e^{i\varphi_{R_0}}}{\Delta m_{K\pi}^2}\ .
}
with $f_+$ and $\xi_{R_0}$ given in~\cite{Epifanov:2007rf,Descotes-Genon:2019bud}.

\section{Narrow-width limit and finite-width effects}
\label{sec:nwl}
\setcounter{equation}{0}

One possible application of the result for the $B\to D M\pi$ decay amplitude derived in the previous section is to regard the three-body decay as a correction to $B\to D V$ in the presence of a finite width of the vector meson, that is, considering the three-body decay as proceeding resonantly through $B\to D\,V (\to M\pi)$.
The integral of the P-wave contribution to the differential decay rate in a window around $k^2=m_V^2$ will be equal to the $B\to D V$ decay rate (for the case of a stable $V$), up to corrections of $\cO(\Gamma_V/m_V)$. This can be shown analytically within the model described in the previous section (see below). In addition, knowing the total width of the vector resonance $V$, one can estimate the corrections to the narrow width limit.

\subsection{Narrow-width limit}

As a first step, one can derive the narrow-width limit of the $\bar B\to D^+M^-\pi^0$ decay
in the single pole approximation, and check that one recovers the known $B\to DV$ result.
We start plugging the model for the LCDA coefficients $B_{n1}^{M\pi}(s)$ of~\Eq{eq:Bn1} into the QCDF amplitude $a_1(D^+ M^-\pi^0)$ to find
\eq{
a_1\big(D^+V^-(\to M^-\pi^0)\big) = 
 \frac{\sqrt{\lambda_{M\pi}(s)}\cos\theta_\pi}{s}  \frac{m_{V}\, f_{V}\, g_{VM\pi}\, e^{i\varphi_{V}}}{\sqrt{2} [m_{V}^2 - s- i\,\sqrt{s}\,\Gamma_{V}(s)]}\; a_1(D^+ V^-)\ ,
}
where it has been assumed that the $M^-\pi^0$ pair arises from the strong decay of a vector resonance $V$, and thus it is purely on a $P$-wave.
The squared of the amplitude is then given by
\eq{
\big|a_1\big(D^+V^-(\to M^-\pi^0)\big)\big|^2  = 
\frac{\lambda_{M\pi}(s)\cos^2\theta_\pi}{2 s^2}
\bigg[
\frac{g_{VM\pi}^2\, f_{V}^2\,m_{V}^2}{(m_{V}^2 - s)^2+ s\,\Gamma_{V}(s)^2}
\bigg]
\; |a_1(D^+ V^-)|^2\ .
}
Using the expression for the strong coupling $g_{VM\pi}$ in terms of the width of the vector resonance given in~\Eq{eq:gRMpi}, one has that
\eqa{
\frac{g_{VM\pi}^2 f_V^2 m_V^2}{(m_V^2 - s)^2 + s \Gamma_V(s)^2} &=&
\frac{48 \pi^2 f_V^2 m_V^2 s^2}{\lambda_{M\pi}^{3/2}(s)}
\,\B(V\to M\pi)
\,\bigg[
\frac1\pi\frac{\sqrt{s}\Gamma_V(s)}{(m_V^2 - s)^2 + s \Gamma_V(s)^2}
\bigg]
\nonumber\\
&&\xlongrightarrow[]{\; \Gamma_V^{\rm tot} \to 0 \;}
\frac{48 \pi^2 f_V^2 m_V^6}{\lambda_{M\pi}^{3/2}(m_V^2)}\,\B(V\to M\pi)\,\delta(s-m_V^2)\ ,
}
where we have used the fact that the term in square brackets goes to a delta function in the limit where the total width of $V$ goes to zero.
Thus, the narrow-width limit of the QCDF amplitude becomes
\eq{
\big|a_1\big(D^+V^-(\to M^-\pi^0)\big)\big|^2
\xlongrightarrow[]{\; \Gamma_V^{\rm tot} \to 0 \;}
\frac{24 \pi^2 f_V^2 m_V^2 \, \cos^2\theta_\pi}{\sqrt{\lambda_{M\pi}(m_V^2)}}
\, |a_1(D^+ V^-)|^2
\,\B(V\to M\pi)\,\delta(k^2-m_V^2)\ ,
}
which leads to the following narrow-width limit for the differential decay rate (c.f.~\Eq{eq:dGamma}),
\eqa{
\frac{d\Gamma(\bar B\to V^-(\to M^-\pi^0))}{ds\,d\cos\theta_\pi}
\xlongrightarrow[]{\; \Gamma_V^{\rm tot} \to 0 \;}&&
G_F^2 \,|V_{ux}^* V_{cb}|^2\ (m_B^2-m_D^2)^2\ |F_0^{B\to D}|^2\, |a_1(D^+ V^-)|^2 \cos^2\theta_\pi
\nonumber\\
&&\times\frac{3 f_V^2 \sqrt{\lambda_{BD}(m_V^2)}}{64 \pi m_B^3}
\,\B(V\to M\pi)\,\delta(s-m_V^2)\ .
}
Now, integrating over the angle $\theta_\pi$ and over the invariant squared mass of the dimeson we have
\eqa{
\Gamma(\bar B\to D^+M^-\pi^0)
&\xlongrightarrow[]{\; \Gamma_V^{\rm tot} \to 0 \;}&
G_F^2 \,|V_{ux}^* V_{cb}|^2\ (m_B^2-m_D^2)^2\ |F_0^{B\to D}|^2\, |a_1(D^+ V^-)|^2 
\nonumber\\
&&\times\ \frac{f_V^2 \sqrt{\lambda_{BD}(m_V^2)}}{32 \pi m_B^3}
\,\B(V\to M\pi)
\nonumber\\
&&=
\frac{\sqrt{\lambda_{BD}(m_V^2)}}{32 \pi m_B^3} \ 2\, |\A(\bar B\to D^+V^-)|^2 \,\B(V\to M\pi)
\nonumber\\
&&= \Gamma(\bar B\to D^+V^-) \ \B(V\to M\pi) \ .
}
Since the differential decay rate in the narrow-width limit contains a delta function, the integral over the invariant mass of the dimeson can be actually be restricted to a narrow window around the mass of the resonance.
This result proves that the narrow-width limit coincides with the two-body decay rate multiplied by the branching ratio of the vector meson to the final dimeson state.
This is just a check that the model has the correct narrow-width limit, however the formalism now allows us to compute corrections to this limit: either finite-width or ``non-resonant" effects.

\subsection{Leading corrections to the narrow-width limit}

We now consider the decay rate integrated in a region around a resonance:
\eq{
\Gamma_{[R]} \equiv \int_{(m_R-\delta)^2}^{(m_R+\delta)^2} ds\, \frac{d\Gamma(\bar B\to D^+M^-\pi^0)}{ds}\ ,
}
where the bin-size $\delta$ is understood to be large enough to contain most of the contribution from the resonance $R$.
Since we have integrated over the angle $\theta_\pi$, the interference terms among different partial waves cancel, and thus:
\eq{
\Gamma_{[R]} = \sum_\ell c_\ell \int_{(m_R-\delta)^2}^{(m_R+\delta)^2} ds\, \frac{\sqrt{\lambda_{BD}(s)\,\lambda_{M\pi}(s)}}{64 (2\pi)^ 3 s m_B^3} |\A^{(\ell)}(s)|^2 = \sum_\ell \Gamma_{[R]}^{(\ell)}
}
with $c_\ell = \int_{-1}^1 dx \, P_\ell(x)^2 = 2/(2\ell+1)$, i.e.\ $c_0 = 2$, $c_1=2/3$ etc.
We now define the ratio:
\eq{
\cW_R^{(\ell)} = \frac{\Gamma_{[R]}^{(\ell)}}{\Gamma_{{[R]}\,,{\rm NWL}}^{(\ell)}}\ ,
}
where $\Gamma_{{[R]}\,,{\rm NWL}}^{(\ell)}$ denotes $\Gamma_{[R]}^{(\ell)}$ in the narrow-width limit, which according to the previous section is given by $\Gamma_{{[R]}\,,{\rm NWL}}^{(\ell)} = \Gamma(\bar B\to D^+R^-) \ \B(R\to M\pi)$.
For example, for the $\rho$ contribution to the $B\to D\pi\pi$ rate we have, neglecting $B_{n1}$ for $n\ge 2$, and using $B_{01}(k^2)=\beta_\pi(k^2) F_\pi(k^2)$:
\eq{
\cW_\rho^{(1)} =
\int_{(m_\rho-\delta)^2}^{(m_\rho+\delta)^2} ds\,
\frac{\lambda^{1/2}_{BD}(s)}{\lambda^{1/2}_{BD}(m_\rho^2)}
\frac{[\beta_\pi(s)]^3\, |F_\pi(s)|^2}{24 \pi^2  f_\rho^2\,\B(\rho\to\pi\pi)}\ .
\label{eq:W_rho}
}
Taking for $F_\pi$ the model specified in ~\Eq{eq:B01pipiModel} gives $\cW_\rho^{(1)} \xlongrightarrow[]{\; \Gamma_\rho^{\rm tot} \to 0 \;} 1$, and thus reproduces the narrow-width limit. This is also obvious from the discussion in the previous section.

\begin{figure}
\centering
\includegraphics[scale=0.87]{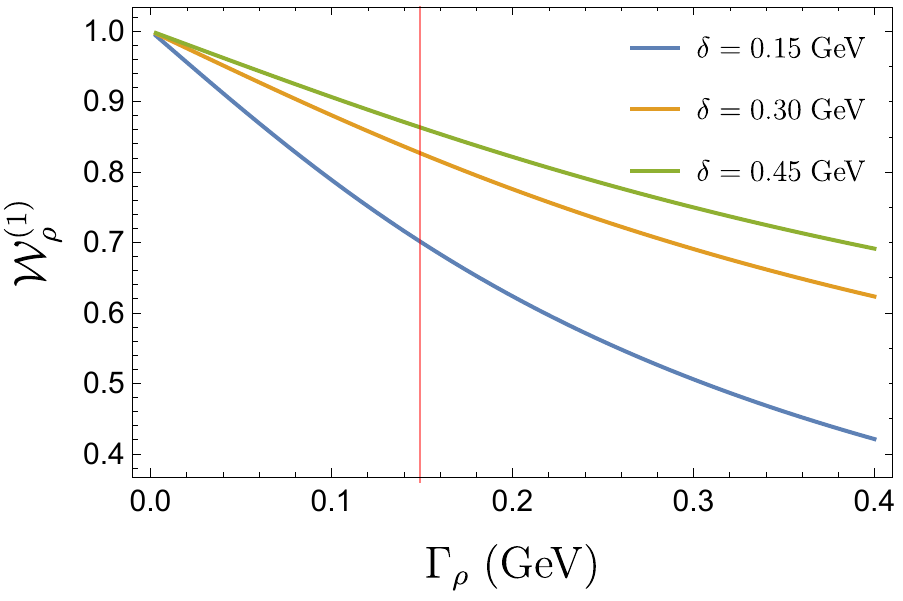}
\hspace{2mm}
\includegraphics[scale=0.87]{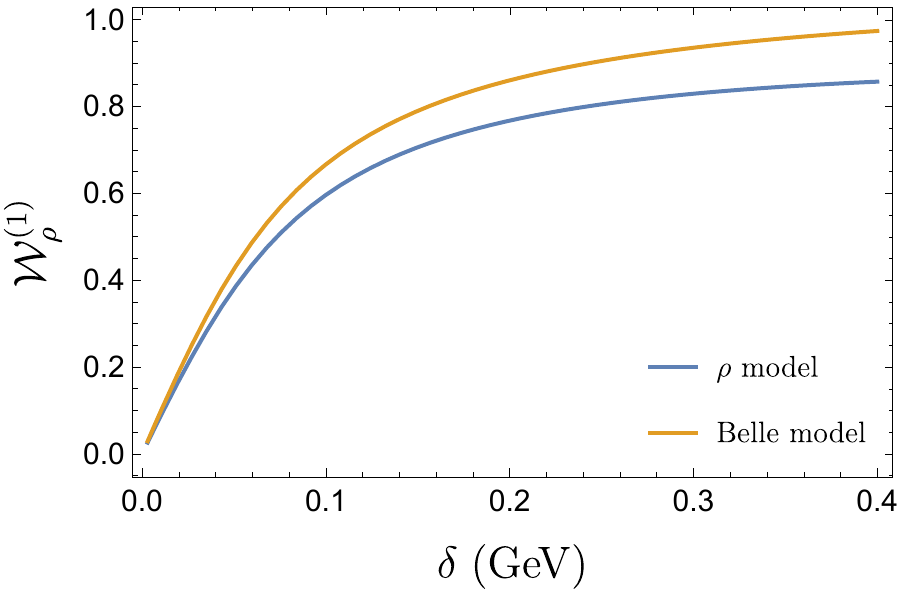}
	\caption{Study of finite-width and bin-size effects on $\cW_\rho$. Left: Corrections to the narrow-width limit of the $\rho$ model in \Eq{eq:W_rho}. The vertical band indicates the physical width $\Gamma_\rho = (149.1\pm0.8)$ MeV \cite{Tanabashi:2018oca}. Right: $\rho$ model as specified in the text and the Belle model \cite{Fujikawa:2008ma} as function of the bin size $\delta$.}
	\label{fig:finw}
\end{figure}

One thing to note from the outset is that the way the resonance model is constructed, whenever we are considering the case that only the resonance $R$ contributes to the quantity $\Gamma_{[R]}^{(\ell)}$, the ratio $\cW_R^{(\ell)}$
does not depend on the matching coefficients in the QCDF amplitude, but only on the properties of the dimeson system (e.g.~\Eq{eq:W_rho}). Nevertheless, it is instructive to consider this case and study the interplay between the width of the single resonance $R$ and the bin-width $\delta$. We consider the ``$\rho$-model'', defined in \Eq{eq:B01pipiModel} keeping only the $\rho$ resonance and leaving its width $\Gamma_\rho$ as a free parameter. In the left panel of Fig.~\ref{fig:finw}, we show the ratio $\cW_\rho^{(1)}$ for this model for different values of $\delta$. The deviation from $\cW_\rho^{(1)}=1$ shows the effect of the finite width and can be understood as the correction to the narrow-width limit. Due to the relatively large width of the $\rho$ meson, we observe an effect of $20\%$ for typical bin sizes of a few times the physical width. In this case, the fact that we have to integrate over a specific bin size makes it more challenging to quantify the finite-width effect. This is different from QCD sum rules for $B\to \pi\pi$ form factors, where this effect can be parametrized by a ``universal" constant $c_R$, defined by $\cW_R^{(\ell)} = 1 + c_R \frac{\Gamma_R}{m_R}$ \cite{Descotes-Genon:2019bud}. We may also study a more realistic model for $F_\pi$ by including also the $\rho'$ and $\rho''$ resonances, for example by considering the model obtained by the Belle Collaboration \cite{Fujikawa:2008ma} (see also \cite{Cheng:2017smj}). We note that this model does not obey the narrow-width limit for the $\rho$ meson. Setting $\Gamma_\rho$ at its physical value, we can study the effects of the $\rho$ width on $\cW_R^{(\ell)}$ depending on the bin size $\delta$. This is illustrated in the right panel of Fig.~\ref{fig:finw}. Compared to the single $\rho$ model, the effect of the heavier resonances in fact slightly reduces the correction to the narrow-width limit. This shows the non-trivial interplay between the different resonances (already when considering only $P$ wave resonances) and underlines the importance of taking such effects into account.

\begin{figure}
\centering
\includegraphics[scale=0.87]{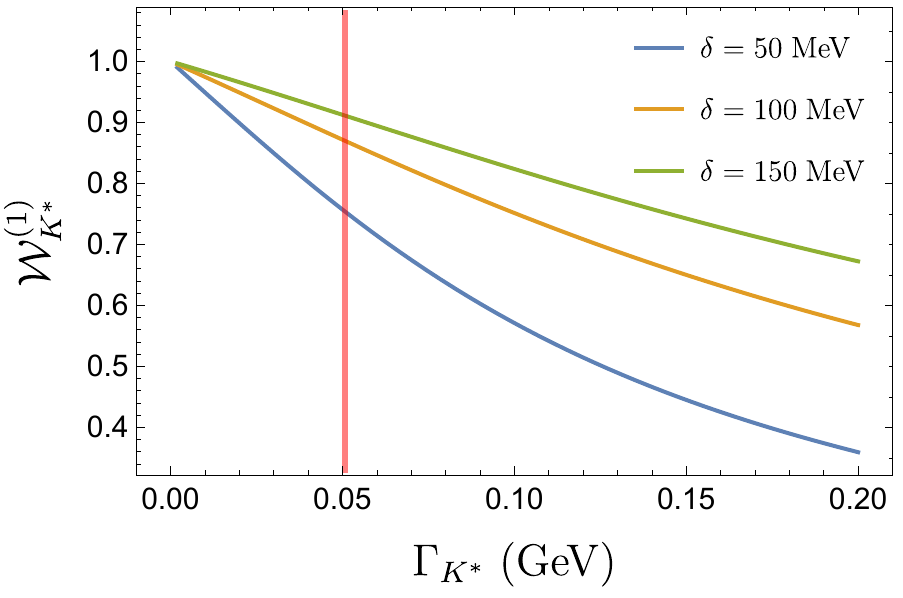}
\hspace{2mm}
\includegraphics[scale=0.87]{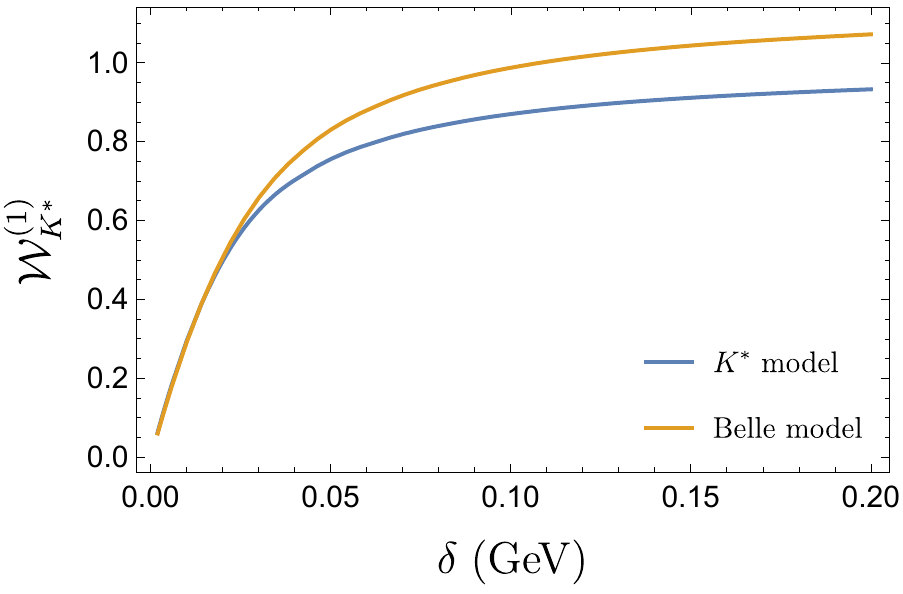}
	\caption{Study of finite-width and bin-size effects on $\cW_{K^*}$. Left: Corrections to the narrow-width limit of the $K^*$ model in \Eq{eq:W_Kstar}. The vertical band indicates the physical width $\Gamma_{K^*} = (50.8\pm0.9)$ MeV \cite{Tanabashi:2018oca}. Right: $K^*$ model and the Belle model \cite{Epifanov:2007rf} as specified in the text as a function of the bin size  $\delta$.}
	\label{fig:finwKstar}
\end{figure}

In the $K\pi$ system, we can perform a similar study considering the $P$-wave $K^*(892)$ resonance. This resonance is much narrower than the $\rho$ and hence the finite width-effects are expected to be somewhat smaller. Explicitly, we find
\eq{
\cW_{K^*}^{(1)} =
\int_{(m_{K^*}-\delta)^2}^{(m_{K^*}+\delta)^2} ds\,
\frac{\lambda^{1/2}_{BD}(s)}{\lambda^{1/2}_{BD}(m_{K^*}^2)}
\frac{[\lambda_{K\pi}(s)]^{3/2}\, |f_+^{K\pi}(s)|^2}{96 \pi^2\,s^3  f_{K^*}^2\,\B({K^*}\to\pi K)}\ .
\label{eq:W_Kstar}
}
Repeating the analysis, we consider the contribution of the $K^*\equiv K^*(892)$ resonance to the form factor $f_+^{K\pi}$, following Eq.~\eqref{eq:B01piKModel} while keeping $\Gamma_{K^*}$ as a free parameter. The finite-width effect is shown in Fig.~\ref{fig:finwKstar}, which is typically around the $10\%$ level. In addition, we may consider a more realistic model containing several resonances. For example, the Belle Collaboration discusses two models obtained by fitting the $\tau\to K_{\rm S} \pi \nu$ decay including both scalar and vector resonances \cite{Epifanov:2007rf} (see also \cite{Descotes-Genon:2019bud}). By definition, $\cW_{K^*}^{(1)}$ only includes the $P$ wave resonances, and we assume a perfect experimental separation of the different partial waves. Therefore, we consider only the Belle model including the vector resonances $K^*(892)$ and $K^*(1410)$. The effect of the bin size $\delta$ for this model is also shown in Fig.~\ref{fig:finwKstar}. We emphasize that besides the effect of heavier $P$ wave resonances, also $S$ wave resonances may still influence or at the least introduce a model dependence in the determination of observables containing a $K^*$. Our simple analysis shows that these effects may be sizeable and should be carefully considered in experimental and theoretical analyses.

\section{Probing higher-order QCD effects}
\label{sec:higherorderQCDF}
\setcounter{equation}{0}

Of interest is also the study of the higher-order QCD effects to which the discussion of the finite-width effects in the previous section was insensitive. To this end it is instructive to look at ratios of differential rates, evaluated at the same value
of $k^2$ but integrated over different ranges of $z\equiv\cos\theta_\pi$. For ease of writing we define the following notation for integrals
\begin{equation}
I[z_1,z_2,f] \equiv \int_{z_1}^{z_2}\! dz \, f(z) \, ,
\end{equation}
and will use again the combination ${\cal G}_n(\mu)$ defined in Eq.~\eqref{eq:CiVi}. 
The ratios that we consider are the following:
\eq{
\mathcal{R}_{M M'}[z_1,z_2;z_1',z_2'](k^2)\equiv\frac{\displaystyle\int_{z_1}^{z_2}\! dz \, \frac{d\Gamma(\bar B \to D^+ M^- \pi^0)}{dk^2 \, dz}}{\displaystyle\int_{z_1^\prime}^{z_2^\prime}\! dz \, \frac{d\Gamma(\bar B \to D^+ M^{\prime-} \pi^0)}{dk^2 \, dz}}\ .
\label{eq:dalitzratios}
}
For $M=M'$, these ratios have the virtue that all prefactors (like the $B\to D$ form factor and CKM elements) cancel, such that:
\eq{
\mathcal{R}_{MM}[z_1,z_2;z_1',z_2'](k^2)   = \frac{\displaystyle\int_{z_1}^{z_2}\! dz \, \left|a_1(D^+ M^-\pi^0)\right|^2}{\displaystyle\int_{z_1^\prime}^{z_2^\prime}\! dz \, \left|a_1(D^+ M^-\pi^0)\right|^2}\ .
}
For the di-pion system alone, $M = M^\prime = \pi$, we will study the effect of higher Gegenbauer moments (i.e.\ $a_2^{\pi\pi}$) and higher-partial waves (i.e.\ $B_{23}$ at $\ell=3$). In addition we will investigate the interplay between $S$ and $P$ wave contributions in the ratio of $K\pi$ final states, $M = M^\prime = K$.

\subsection{Di-pion system}

In the isospin limit, only odd partial waves contribute to the dipion wave function. Thus the squared amplitude $|a_1(D\pi\pi)|^2$ contains no interference terms between even and odd partial waves, and it is symmetric under $z\to -z$. As a result, the ratios $\mathcal{R}_{\pi\pi}$ satisfy relations such as
\eq{
\mathcal{R}_{\pi\pi}[0,z;-z,z](k^2)=\mathcal{R}_{\pi\pi}[-z,0;-z,z](k^2)=\frac12\ .
}
A simple corollary is that the ``forward-backward" asymmetry vanishes,
\eq{
A^{\pi\pi}_{\rm FB}(k^2) = \mathcal{R}_{\pi\pi}[0,1;-1,1](k^2)-\mathcal{R}_{\pi\pi}[-1,0;-1,1](k^2) = 0\ .
}
These kind of observables thus constitute a simple test for isospin-violating corrections.

In general, and keeping terms up to $n=2$ in the Gegenbauer expansion, we have
\eq{
\mathcal{R}_{\pi\pi}[z_1,z_2,z_1',z_2'](k^2)
= \frac{\displaystyle\int_{z_1}^{z_2}\! dz \, \left| \alpha_0^{\pi\pi}(k^2,\theta_\pi) \, {\cal G}_0(\mu_b) + \alpha_2^{\pi\pi}(k^2,\theta_\pi) \, {\cal G}_2(\mu_b) \right|^2}{\displaystyle\int_{z_1^\prime}^{z_2^\prime}\! dz \, \left|\alpha_0^{\pi\pi}(k^2,\theta_\pi) \, {\cal G}_0(\mu_b) + \alpha_2^{\pi\pi}(k^2,\theta_\pi) \, {\cal G}_2(\mu_b)\right|^2} \ ,
\label{eq:dipionratio}
}
and expanding Eq.~(\ref{eq:dipionratio}) for small ${\cal G}_2$ (respectively small $\alpha_2^{\pi\pi}$) one finds
\eqa{
\mathcal{R}_{\pi\pi}[z_1,z_2,z_1',z_2'](k^2) &=& \frac{\displaystyle I[z_1,z_2,|\alpha_0^{\pi\pi}|^2]}{\displaystyle I[z_1^\prime,z_2^\prime,|\alpha_0^{\pi\pi}|^2]} + \frac{\displaystyle 2 \, {\rm{Re}}\left(I[z_1,z_2,\alpha_0^{\pi\pi}\alpha_2^{\pi\pi \, \ast}] \, I[z_1^\prime,z_2^\prime,|\alpha_0^{\pi\pi}|^2] \, {\cal G}_0 \, {\cal G}_2^\ast\right)}{\displaystyle \left|{\cal G}_0(\mu_b)\right|^2 (I[z_1^\prime,z_2^\prime,|\alpha_0^{\pi\pi}|^2])^2}
\nonumber \\[3mm]
&& - \frac{\displaystyle 2 \, {\rm{Re}}\left(I[z_1^\prime,z_2^\prime,\alpha_0^{\pi\pi}\alpha_2^{\pi\pi \, \ast}] \, I[z_1,z_2,|\alpha_0^{\pi\pi}|^2] \, {\cal G}_0 \, {\cal G}_2^\ast\right)}{\displaystyle \left|{\cal G}_0(\mu_b)\right|^2 (I[z_1^\prime,z_2^\prime,|\alpha_0^{\pi\pi}|^2])^2}\ .
\label{eq:dipionsmallalpha2}
}
The leading part can be rewritten as $I[z_1,z_2,(P_1(z))^2]/I[z_1^\prime,z_2^\prime,(P_1(z))^2]$ and hence only depends on angular variables. The dependence of the correction term on quantities of interest ($k^2$, higher Gegenbauer moments, QCD corrections) is, however, 
quite involved. It is therefore instructive to study the effect of higher partial waves, for which we substitute
\begin{align}
\alpha_0^{\pi\pi}(k^2,\theta_\pi) &= B^{\pi\pi}_{01}(k^2)\, P_1(z) \, , \nonumber \\[0.3em] \alpha_2^{\pi\pi}(k^2,\theta_\pi) &= B^{\pi\pi}_{21}(k^2)\, P_1(z) + B^{\pi\pi}_{23}(k^2)\, P_3(z), 
\end{align}
into (\ref{eq:dipionratio}), assume $P$-wave dominance and thus treat $B_{23}$ at $\ell=3$ as a correction. Eq.~(\ref{eq:dipionratio}) then becomes
\begin{align}
&\mathcal{R}_{\pi\pi}[z_1,z_2,z_1',z_2'](k^2) =\frac{\displaystyle I[z_1,z_2,P_1^2]}{I[z_1^\prime,z_2^\prime,P_1^2]} + \frac{\displaystyle I[z_1,z_2, P_1 \, P_3] \, I[z_1^\prime,z_2^\prime,P_1^2] - I[z_1^\prime,z_2^\prime, P_1 \, P_3] \, I[z_1,z_2,P_1^2]}{\left(I[z_1^\prime,z_2^\prime,P_1^2]\right)^2} \nonumber \\[0.3em]
&\qquad \times \frac{\displaystyle 2 \, {\rm{Re}}\left(B^{\pi\pi}_{01}(k^2) \, B^{\pi\pi \, \ast}_{23}(k^2) \, {\cal G}_0(\mu_b) \, {\cal G}_2^\ast(\mu_b) \right) + 2 \, {\rm{Re}}\left(B^{\pi\pi}_{21}(k^2) \, B^{\pi\pi \, \ast}_{23}(k^2) \right) |{\cal G}_2(\mu_b)|^2}{\displaystyle \left|B^{\pi\pi}_{01}(k^2) \, {\cal G}_0(\mu_b) + B^{\pi\pi}_{21}(k^2) \, {\cal G}_2(\mu_b) \right|^2} \, .
\label{eq:dipionPwavedom}
\end{align}
We observe that the leading term is again given by (the same) angular integrals. However, the angular dependence now factorises {\emph{at each order}}, and hence the sensitivity of the correction term to $k^2$, higher Gegenbauer moments and the NNLO QCD corrections is more transparent. The actual size of the correction depends on the value of $k^2$ and the bins in $z=\cos\theta_\pi$ under consideration. For instance,
\begin{align}
\mathcal{R}_{\pi\pi}[-1/2,1/2,-1,1](k^2) & \approx  \frac{1}{8} -0.28 \,  {\rm{Re}}\left[
\frac{B^{\pi\pi}_{23}(k^2)}{B^{\pi\pi}_{01}(k^2)} 
\,
\frac{{\cal G}_2(\mu_b)}{{\cal G}_0(\mu_b)}
\right] \ ,
\end{align}
where we have again used that ${\cal G}_2/{\cal G}_0 \simeq 0.4\%$ is small.

\subsection{$K \pi$ system}

The main differences between the $K \pi$ and $\pi \pi$ final states are the presence of the first Gegenbauer moment $\alpha_1^{K\pi}$ and the appearance of partial waves of even $\ell$. The ratio in this case reads
\eq{
\mathcal{R}_{KK}[z_1,z_2,z_1',z_2'](k^2)\!
= \frac{\displaystyle\int_{z_1}^{z_2}\!\!\! dz \! \left| \alpha_0^{K\pi}(k^2,\theta_\pi) \, {\cal G}_0(\mu_b) + \alpha_1^{K\pi}(k^2,\theta_\pi) \, {\cal G}_1(\mu_b) + \alpha_2^{K\pi}(k^2,\theta_\pi) \, {\cal G}_2(\mu_b) \right|^2}{\displaystyle\int_{z_1^\prime}^{z_2^\prime}\!\!\! dz \! \left|\alpha_0^{K\pi}(k^2,\theta_\pi) \, {\cal G}_0(\mu_b) + \alpha_1^{K\pi}(k^2,\theta_\pi) \, {\cal G}_1(\mu_b) +\alpha_2^{K\pi}(k^2,\theta_\pi) \, {\cal G}_2(\mu_b)\right|^2}.
\label{eq:dikaonratio}
}
The general structure of this expression is rather complicated, but a simple observable that can be studied in this case is the forward-backward asymmetry
\eq{
A^{K\pi}_{\rm FB}(k^2) = \mathcal{R}_{K\pi}[0,1;-1,1](k^2)-\mathcal{R}_{K\pi}[-1,0;-1,1](k^2) \ .
}
In this case, the numerator is sensitive to interference of even and odd partial waves, whereas in the denominator all partial waves are separated. If we furthermore expand in the small quantities
${\cal G}_1$ and ${\cal G}_2$ and work under the assumption that $S$ and $P$-wave dominate, we arrive at the simplified expression
\begin{align}
A^{K\pi}_{\rm FB}(k^2) \simeq & \frac{2 {\rm Re}(B_{00} B^*_{01})}{2 |B_{00}|^2+2/3 |B_{01}|^2}  \nonumber \\[0.5em]
+& \frac{2 {\rm Re}\left[\left(2 (B^*_{00})^2 - 2/3 (B^*_{01})^2\right) {\cal G}_0^* \left(B_{00} (B_{11} {\cal G}_1 + B_{21} {\cal G}_2)-B_{01} (B_{10} {\cal G}_1 + B_{20} {\cal G}_2)\right)\right]}{\left|{\cal G}_0\right|^2\left(2 |B_{00}|^2+2/3 |B_{01}|^2\right)^2}\ ,
\label{eq:AFBKpi}
\end{align}
where we have omitted the $K\pi$ superscripts on the coefficients $B_{n\ell}^{K\pi}$.
To leading order in $\alpha_s$, $A^{K\pi}_{\rm FB}$ is proportional to the real part of $B_{00} B^*_{01}\sim \Delta m_{K\pi}^2\,f_0^{K\pi} f_+^{K\pi\,*}$, and vanishes in the limit $m_K=m_\pi$.
This corresponds to the vanishing of $A^{\pi\pi}_{\rm FB}$ in the isospin limit, and in fact this is true to all orders if $B_{n0}$ is proportional to $\Delta m_{K\pi}^2$ for all $n$.
Starting at NLO, one starts probing the higher Gegenbauer coefficients $B_{11}$, $B_{20}$ and $B_{21}$.
It is straightforward to generalise the expression~(\ref{eq:AFBKpi}) to the case of including higher partial waves: One separately expands numerator and denominator of~(\ref{eq:dikaonratio}), performs the angular integrations and subsequently expands the entire expression to linear order in ${\cal G}_1$ and ${\cal G}_2$.

\section{Discussion and conclusions}
\label{sec:conclusions}
\setcounter{equation}{0}

After two decades of intense work on the theory of two-body non-leptonic $B$ decays, leading-power factorization has been established at NNLO in QCD (see~\cite{Bell:2020qus} for the most recent work). The extension to \emph{three-body} decays has been pursued more recently, with their share of phenomenological virtues and theoretical complications. Here we have studied what might be the simplest of such three-body decays from the point of view of factorization: the heavy-to-heavy $B\to D M \pi$ decays ($M=\pi,K$) in the kinematic region where the $M\pi$ dimeson system has small invariant mass compared to the available energy. The result is a factorized amplitude, given in~Eqs.~(\ref{eq:A(B->DMpi)}) and~(\ref{eq:a1NNLO}), which includes NNLO $\alpha_s(m_b)$ corrections, and which generalizes in a simple way the formula for the two-body decay.

One particularity of the three-body decay amplitude is that it reproduces analytically the corresponding well-known amplitudes for the quasi-two body decays $B\to D\rho$ and $B\to D K^*$ in the case where the decay occurs through an intermediate infinitely-narrow resonance. This allows to study corrections to the narrow-width limit and to interpret measurements of quasi-two-body decays. We have introduced a quantity, $\cW^{(\ell)}_R$, which quantifies the departure from the narrow-width limit of a particular measurement of a decay rate in the region around a resonance. We find that both for the $\rho$ and for the $K^*$, a typical value for this quantity is $\cW^{(1)}_R\sim 0.8$, meaning that interpreting these measurements in the narrow-width limit, the missing finite-width effect is of the order of $20\%$.

Beyond the study of finite-width effects in quasi-two-body decays, we have considered ratios of Dalitz-plot bins with equal values for the dimeson invariant mass, for both $B\to D\pi\pi$ and $B\to D K \pi$. These ratios are interesting because many quantities cancel out (such as form factors and CKM elements), and probe directly a product of higher-order $\alpha_s$ corrections and higher Gegenbauer moments of the dimeson LCDAs. One particular example is the forward-backward asymmetry. We believe that these sort of observables might be interesting phenomenologically once experimental data accumulates.

\section*{Acknowledgements}

We thank Pablo Roig, Martin Savior, Kim Smith and Sergi Gonz\`alez-Sol\'is for useful discussions and correspondence.
J.V. acknowledges funding from the European Union's Horizon 2020 research and innovation program under the Marie Sklodowska-Curie grant agreement No 700525, `NIOBE' and from the Spanish MINECO through the “Ramon y Cajal” program RYC-2017-21870. The work of TH was supported in part by the Deutsche Forschungsgemeinschaft (DFG, German Research Foundation) under grant  396021762 - TRR 257 ``Particle Physics Phenomenology after the Higgs Discovery''. The work of KKV is supported by the DFG Sonderforschungsbereich/Transregio 110 ``Symmetries and the Emergence of Structure in QCD''. We would like to thank to the Mainz Institute for Theoretical Physics (MITP) of the Cluster of Excellence PRISMA+ (Project ID 39083149) for its hospitality and support during the 2019 MITP workshop ``Future Challenges in Non-Leptonic B-Decays".

 \newpage


\end{document}